\documentclass[prd, twocolumn, nofootinbib, floatfix]{revtex4-1}

\usepackage{amsmath}
\usepackage{graphicx}
\usepackage{dcolumn}
\usepackage{bm}
\usepackage{epsfig}
\usepackage{amssymb,latexsym,mathrsfs}
\usepackage{graphicx}
\usepackage{color}
\usepackage{hyperref}

\hypersetup{
    colorlinks=true,
    linkcolor=red,
    citecolor=blue,
} 

\newcommand{\be}{\begin{equation}}
\newcommand{\ee}{\end{equation}}
\newcommand{\bs}{\begin{split}} 
\newcommand{\bea}{\begin{eqnarray}}
\newcommand{\eea}{\end{eqnarray}}

\newcommand{\al}{\alpha} 
\newcommand{\kap}{\kappa} 
\newcommand{\kp}{\kappa}
\newcommand{\eps}{\epsilon}

\begin{document}

\title{Eternal and Evanescent Black Holes: It's All Done With Mirrors} 
\author{Michael R.R. Good${}^{1,2}$}
\author{Eric V.\ Linder${}^{2,3}$} 
\affiliation{${}^1$Physics Department, School of Science and Technology, Nazarbayev University, Astana, 
Kazakhstan\\
${}^2$Energetic Cosmos Laboratory, Nazarbayev University, Astana, 
Kazakhstan\\ 
${}^3$Berkeley Center for Cosmological Physics \& Berkeley Lab, 
University of California, Berkeley, CA 94720, USA} 

\begin{abstract}
The analogy between black hole radiation and accelerating mirror radiation (the dynamical 
Casimir effect) is particularly strong for mirror trajectories giving rise 
to a constant thermal flux of particles. We present new ways to achieve such 
thermal plateaus, and customize their finite, semi-infinite, and eternal 
presence, corresponding to forming/collapsing, complete-evaporation/remnants, and eternal black holes. 
We find simple expressions for the energy flux in terms of the mirror rapidity 
as a function of proper time and null time. 
\end{abstract} 

\date{\today} 

\maketitle

\section{Introduction} 

The understanding of black hole evaporation \cite{Hawking:1974sw} in the context of thermal emission with temperature 
\be T = \frac{\kappa}{2\pi}\ , \label{temperature}\ee
where $\kap$ is the surface gravity, has been greatly facilitated by  investigations into flat spacetime acceleration phenomena like the Unruh effect \cite{Unruh:1976db} or the DeWitt-Davies-Fulling effect \cite{DeWitt:1975ys, Davies:1976hi, Davies:1977yv}. This can be viewed as an addendum to the Principle of 
Equivalence relation between gravity and acceleration. 

The Unruh effect occurs with hyperbolic motion of constant proper acceleration $\alpha = \kappa$, i.e.\ without time dependence $\alpha(t)$, which is a Lorentz invariant in all references frames. 

The DeWitt-Davies-Fulling effect is the moving mirror model of the dynamical Casimir effect \cite{Moore:1970} where the perfectly reflecting boundaries move along trajectories that have \textit{dynamic} acceleration, $\alpha = \alpha(t)$.  

The flat spacetime temperature from Eq.~(\ref{temperature}) results from a constant $\kappa$ that has been demoted to \textit{only} a parameter of the accelerating mirror system, which sets the temperature scale of the problem. Interestingly, and in stark contrast to the Unruh effect, it does not directly represent the proper acceleration of the mirror.  In fact, the particular clock used determines the form of the proper acceleration, $\alpha(t)$. 

In this article we explore the expression for thermal emission in terms 
of different time coordinates, determine simple relations and exact 
solutions for the forms of acceleration that give rise to thermal 
emission, and study the rise and possible end state of such emission. 
This offers the possibility of shedding some light on analogous 
processes for black holes, in particular evaporation and remnants, 
or eternality. 

In Sec.~\ref{sec:time} we investigate thermality as it appears in 
null time and proper time, finding great simplicity of 
expression in the latter case. Section~\ref{sec:zdot} then returns to 
the usual picture in terms of mirror time and shows the close relation. 
We apply these insights to known solutions in Sec.~\ref{sec:known}, to 
asymptotic solutions in Sec.~\ref{sec:asymp}, 
and then create bespoke solutions in Sec.~\ref{sec:bespoke}. 
We summarize and conclude in Sec.~\ref{sec:concl}.

\section{Energy Flux, Acceleration, and Time}\label{sec:time} 

We begin by investigating the energy flux of the Davies-Fulling renormalized stress-energy \cite{Davies:1976hi,Davies:1977yv} (see also e.g.\ \cite{Walker:1984vj,walkerdavies}) from a moving mirror as a function of the 
retarded-time clock or null-time clock $u$ of the observer at infinity. This is all in the context of Minkowski space with one time dimension and one space dimension $z$. The energy flux can be written in terms of the rapidity, $\eta(u) = (1/2)\ln p'(u)$, where $p(u)$ is the ray tracing function, as 
\be 12\pi F(u) = \eta'(u)^2 - \eta''(u)\ , \ee 
(see e.g.\ Wilczek \cite{Wilczek:1993jn}). 
  
This can be further simplified by using the 
acceleration $\al(u)$. To find $\al(u)$, note that $\al(\tau)=d\eta/d\tau$, and that with the null coordinates $u=t-z$ 
and $v=t+z$, we can write $d\tau^2 = du dv$, where $v=p(u)$. Thus one has
\be \alpha(u) = \eta'(u)\,e^{-\eta(u)}=-\frac{d}{du}\,e^{-\eta(u)}\ . 
\ee 
The final formula for the flux is simply written in terms of the dynamics, 
\be 12 \pi F(u) = -\alpha'(u)\,e^{\eta(u)}=e^{\eta}\,\frac{d^2}{du^2}\,e^{-\eta}\ . 
\ee 

One can readily verify that to obtain a constant energy flux with Planckian temperature, Eq.~(\ref{temperature}), the rapidity form $\eta(u) = -\kappa u/2$ works (see e.g.\ \cite{Good:2016atu,Carlitz:1986nh}). 
The constant energy flux is 
\be  F = \frac{\kappa^2}{48 \pi}\ .\label{constantenergyflux}\ee
This means that the proper acceleration is exponential, 
\be \alpha(u) = -\frac{\kappa}{2} e^{\kappa u/2}\ , \quad \Rightarrow \quad T = \frac{\kappa}{2\pi}\ . \label{expacc} \ee
Therefore, exponential acceleration as measured using $u$ gives rise to temperature, Eq.~(\ref{temperature}), in the dynamical Casimir effect.  

However, the form is quite different using the clock of a traveler moving alongside the moving mirror.  The acceleration as a function of proper time $\tau$, i.e.\ $\alpha(\tau)$, can be found simply by converting the null-time, $u$, expression to the proper time expression by means of 
\be 
\frac{d \tau}{du} = e^\eta \ . 
\ee 
We find the simple formula,  
\be 12\pi F(\tau) = -\eta''(\tau) e^{2\eta(\tau)} = - \alpha'(\tau) e^{2\eta(\tau)}\ . \label{eq:ftau}\ee 

We can now immediately understand what mirror solutions give thermal flux plateaus. 
For one, rapidity $\eta(\tau) = \ln(\kappa\tau/2)$ gives constant energy flux for all times: 
$48\pi F = \kappa^2$. That is, Eq.~(\ref{constantenergyflux}) results from scale-independent acceleration:
\be \alpha(\tau) = \tau^{-1}\ , \quad \Rightarrow \quad T = \frac{\kappa}{2\pi}\ . \ee 
Therefore, scale-independent acceleration, inversely proportional to proper time $\tau$, 
gives rise to temperature in the dynamical Casimir effect. 

In fact, we can be more general and complete. From Eq.~(\ref{eq:ftau}) the condition that $F=\,$constant is 
simply $\eta''\,e^{2\eta}=\,$constant. Writing this as the  differential equation  
\be 
\eta'''+2\eta'\eta''=0\ , 
\ee 
we expect three solutions. Using $\al=\eta'$, we have 
\be 
\al''+(\al^2)'=(\al'+\al^2)'=0\ , 
\ee 
or 
\be 
\al'+\al^2+C=0\ , 
\ee 
where $C$ is a constant. 
The solutions for the acceleration can easily be seen to be 
\be 
\al=\frac{1}{\tau},\quad -\frac{\kap}{2}\tan\frac{\kap\tau}{2},\quad 
-\frac{\kap}{2}\tanh\frac{\kap\tau}{2}\ , 
\ee 
or in terms of rapidity  
\be 
\eta=\ln\frac{\kap\tau}{2},\quad \ln\cos\frac{\kap\tau}{2}, 
\quad \ln\cosh\frac{\kap\tau}{2}\ .  \label{eq:plateta}
\ee 

The first solution is the scale independent acceleration, giving eternal thermality. 
In Appendix~\ref{sec:apxcw} we show this is none other than the Carlitz-Willey 
solution \cite{Carlitz:1986nh}, written in a 
far simpler form without product logs \cite{Good:2012cp,Good:2013lca}, i.e.\ without the Lambert $W$ function.

\section{Energy Flux and Mirror Trajectory} \label{sec:zdot} 

Let us investigate the first general plateau solution further. To see what mirror 
trajectories have the property of thermal flux asymptotically, we want to convert from 
acceleration as a function of 
proper time to mirror position $z$ as a function of coordinate time $t$. Recall that 
$\alpha(\tau)=1/\tau$ corresponds to $\eta(\tau)=\ln (\kp\tau/2)$. We can relate this 
to the Lorentz factor $\gamma$ by $\cosh\eta=\gamma$ (see Appendix~\ref{sec:apxtsfm}). Asymptotically, 
near $\tau\approx0$ 
we then have $\gamma\approx -1/(\kap\tau)$. Using $\gamma\,d\tau=dt$ we find 
$\tau\sim e^{-\kp t}$ and so $\gamma=C\,e^{\kp t}$, where $C$ is a constant. 

Finally, using $\gamma=(1-\dot z^2)^{-1/2}$ we derive the asymptotic behavior 
\be 
\dot z\to -1+C\,e^{-2\kp t}\ . 
\ee 
This describes a mirror asymptotically approaching the speed of light, but more importantly 
with an exponential accession to the speed of light. 
The particular value of $C$ is not crucial to the thermal flux plateau but it is 
convenient to choose $C=2$ to allow $\dot z$ to run over the full range $[-1,+1]$. 
In fact, the 2 in the exponent just comes from the 2 inside the logarithm in $\eta$, 
and so is also somewhat arbitrary (though it is needed to give the amplitude of the 
thermal plateau). The essential factor is the exponential in coordinate time approach to the 
speed of light, or equivalently the reciprocal proper time proportionality of the 
acceleration. 

This thermal behavior has 
tight connections to the analog case of black hole radiation. 
The thermal behavior of a trajectory that at late times has 
$\dot z\to -1+C\,e^{-2\kp t}$, i.e.\ $z\to -t-[C/(2\kp)]e^{-2\kp t}$, 
can be related, through a conformal transformation from Minkowski 
spacetime to Schwarzschild spacetime, to timelike curves crossing 
the Schwarzschild black hole horizon, $r\sim -t-A\,e^{t/(2M)}$ 
\cite{Davies:1977yv,mtw}. (See Eqs.~2.2 and 2.3 of \cite{Davies:1977yv}.) 
Here $M$ is the black hole mass, 
with temperature $T=1/(8\pi M)=\kp/(2\pi)$. 

One can also obtain the asymptotic condition for thermality without reference to 
proper time. 
Consider the general expression for the energy flux in terms of the 
mirror trajectory $z(t)$, or velocity $\dot z$: 
\be 
F=-\frac{1}{12\pi}\frac{\dddot z (1-\dot z^2)+3\dot z \ddot z^2}{(1-\dot z)^4 
(1+\dot z)^2} \ . \label{eq:flux} 
\ee 

In the asymptotically static case, all derivatives $d^n z/dt^n\to0$ and 
so we expect the energy flux to vanish at late times. This will not give 
a thermal plateau. 

In the asymptotically drifting case, suppose the drift velocity is not 
very close to the speed of light. Then the denominator of Eq.~(\ref{eq:flux}) 
just goes to some constant and the numerator involves only $\ddot z$ 
and $\dddot z$ terms. But to reach an asymptotically drifting state 
these must vanish at late times and so again the flux vanishes and 
we do not obtain a thermal plateau. 

Therefore we are left with the case where $\dot z=-1+f(t)$, where $f\to0$ 
asymptotically. 
For $f\ll1$, the energy flux becomes 
\be 
F\approx \frac{1}{48\pi}\left[\frac{3}{4}\left(\frac{\dot f}{f}\right)^2-\frac{1}{2}\left(\frac{\ddot f}{f}\right)\right] \ . \label{eq:fplatf} 
\ee 
To obtain a plateau, the quantity in brackets must be time independent 
during that period. This only occurs for the form $f\propto e^{-m\kp t}$. 
Thus we are basically looking at mirror trajectories with 
\be 
\dot z\to -1+{\mathcal O}(e^{-m\kap t}) \ . \label{eq:zexp} 
\ee

\section{Approach to Thermality} \label{sec:known} 

Having established the conditions for pure thermality, in this section we examine the approach to thermality by considering the dynamics of five different known mirror solutions: Arcx and Darcx \cite{Good:2013lca}, Omex and Domex \cite{Good:2016oey,Good:2016atu}, and a null-self-dual case \cite{paper1}. 
An approach to thermal emission is connected to exponential acceleration in either time or space.  
In all of these cases, the asymptotic behavior is of the form of Eq.~(\ref{eq:zexp}), as 
explored in the previous section and later in more detail in Sec.~\ref{sec:asymp}.

\subsection{Exponential Acceleration in Time} 

\subsubsection{Arcx} \label{sec:arcx}

The simplest example of a mirror that approaches a constant energy flux and temperature is that of Arcx (which stands for ARC-hyperbolic sine of an eXponential)\cite{Good:2013lca}, whose trajectory $z(t)$ is 
\be z(t) = -\frac{1}{\kappa}\sinh^{-1}\frac{e^{\kappa t}}{2}.\ee
The special nature of this trajectory is that its proper acceleration is exponential in coordinate time $t$, 
\be \alpha(t) = -\frac{\kappa}{2} e^{\kappa t},\label{expint}\ee
for all times $t$. The mirror is asymptotically null with maximum asymptotic speed $\dot z \to 1$, and the particle beta Bogolyubov coefficients are easy to solve (see \cite{Good:2013lca}). 

The relationship between coordinate time and proper time is found using the inverse, $t(\tau)$, of 
\be \tau(t) = \int \frac{dt}{\gamma(t)} = - \frac{1}{\kappa} \sinh^{-1}(2 e^{-\kappa t})\ . \ee
Then the trajectory written in terms of proper time is 
\be z(\tau) = \frac{1}{\kappa}\sinh^{-1}\textrm{csch} \; \kappa \tau\ , \ee
and the acceleration of Eq.~(\ref{expint}) becomes,
\be \alpha(\tau) = \kappa\,\textrm{csch}\;\kappa \tau\ , \label{benito}\ee
which is in agreement with \cite{BENITO}. Near $\tau = 0$, the acceleration scales as
\be \alpha(\tau) = \frac{1}{\tau} + \mathcal{O}(\tau), \ee
whose scale independence characterizes the emergent thermal plateau in energy flux that lasts for all late times, $\tau \to 0$ (i.e.\ $t\to\infty$), since $-\infty < \tau < 0$ due to the horizon. One can readily verify 
as well that the asymptotic velocity is $\dot z\to -1+2e^{-2\kap t}$.  For further investigations of this interesting and prototypical moving mirror trajectory see e.g. Benito \cite{BENITO}, Good-Anderson-Evans \cite{Good:2013lca}, and Hotta-Shino-Yoshimura \cite{Hotta:1994ha}.

\subsubsection{Darcx} \label{sec:darcx}

The drifting version of Arcx with maximum speed $v$ has the trajectory \cite{Good:2013lca}, 
\be z(t) = -\frac{v}{\kappa}\sinh^{-1}\frac{e^{\kappa t}}{2}\ , \ee
and acceleration $\alpha(t)$,
\be \alpha(t) =-\frac{4 \kappa  v e^{\kappa  t}}{\left[4+\left(1-v^2\right) e^{2 \kappa  t}\right]^{3/2}}\ .\ee
One can find the proper time as a function of mirror time, $\tau(t)$, but inversion proves difficult.  Instead, it proves instructive to work to first order in large $\gamma \equiv (1-v^2)^{-1/2}$. The result is
\be \tau(t) = \frac{\ln 2}{2 \gamma  \kappa }-\frac{1}{\kappa} \sinh ^{-1}\left(2 e^{-\kappa t}\right) + \mathcal{O}(\gamma^{-1})\ , \ee
which can be inverted to first order,
\be t(\tau) = -\frac{1}{\kappa}\ln\left[ \frac{1}{2} \sinh \left(\frac{\ln 2}{2\gamma} - \kappa  \tau\right)\right]\ . \ee 

Substitution into the acceleration, $\alpha(t) \to \alpha(\tau)$, and an expansion for large $\gamma$ gives to first order 
\be \alpha(\tau) = \kappa\, \textrm{csch}(\kappa \tau) + \frac{\kappa \ln 2}{2\gamma}\,\frac{\cosh \kappa \tau}{\sinh^2 \kappa \tau} + \mathcal{O}(\gamma^{-1})\ . \ee
One sees the leading order term is Arcx's proper acceleration, Eq.~(\ref{benito}). A subsequent expansion around $\tau = 0$, the position of the `residual horizon', gives
\be \alpha(\tau) = \frac{\ln 2}{2\gamma \kappa \tau^2} + \frac{1}{\tau} + \frac{\kappa \ln 2}{12\gamma} + \mathcal{O}(\tau)\ . \ee
Now we can clearly see the conditions for a thermal plateau. If the  first term above is much smaller than the second term, i.e.\  
\be \gamma \kappa \tau \gg 1\ , \ee
and the second term is much bigger than the third term, i.e.\ 
\be \frac{\gamma}{\kappa \tau} \gg 1\ , \ee 
then the second term, $\tau^{-1}$, dominates.  

Taken together, if the plateau time $\tau$ and the fixed parameters $\kappa$ and $\gamma$ are such that
\be \frac{\gamma}{\kappa} \gg \tau \gg \frac{1}{\kappa\gamma}\ , \ee
then the acceleration scales as (independent of $\kp$) 
\be \alpha(\tau) = \frac{1}{\tau}\ , \ee
and an evanescent thermal plateau will be present. The gravitational analog is of a forming black hole that begins to non-thermally radiate, eventually equilibrating to emit thermally for a finite 
lifetime, and when the evaporation process is finished, a remnant remains.  

In this drifting case\footnote{For an alternate drifting case that has a pulse of energy before settling down to thermal emission see \cite{Good:2015nja} or for one that accelerates to the speed of light \cite{Good:2016yht} with zero asymptotic acceleration. }, it is important to recognize there is no horizon, and so $-\infty < \tau < \infty$.  At late times, $\tau \to \infty$ (instead of $\tau \to 0$ for Arcx) and therefore $\alpha(\tau) = \tau^{-1} \to 0$, signaling the coasting end state, where evaporation has completely stopped, i.e.\ the plateau has long fallen off because $\tau$ eventually exceeds the fixed value of $\gamma/\kappa$.  

We explore these conditions in further generality both analytically and numerically 
in Sec.~\ref{sec:asymp}.

\subsection{Exponential Acceleration in Space} 

When the form of the proper acceleration of a mirror in proper time, $\alpha(\tau)$, is intractable 
or very challenging to obtain it is useful to consider the proper acceleration in terms of space.  

\subsubsection{Black Mirror} 

The black mirror trajectory (or Omex) \cite{Good:2016oey} has particle creation that corresponds exactly to a collapsing null shell in curved spacetime. Its spacetime trajectory $z(t)$ is given by 
\be z(t)=-t- \frac{1}{2\kappa} W[2e^{-2\kappa t}]\ , \ee
Where $W$ is the Lambert $W$ function or product log.  
Note that since $W(x\ll1)\approx x$, then $\dot z$ follows the 
asymptotic form of Eq.~(\ref{eq:zexp}). 
The timespace trajectory, $t(x)$, is given by 
\be t(z) = - z - \frac{1}{\kappa}e^{2\kappa z}\ . \ee 

It is easy to find a form of the proper acceleration as a function of space without the product log because of this simple timespace trajectory.  One uses 
\be \alpha(z) = \frac{d}{dz} \gamma(z)\ ,\ee
where the usual Lorentz factor is simply expressed as a function of space, rather than time, 
\be \gamma(z) =\frac{t'(z)}{\sqrt{(t'(z))^2 -1}}\ . \ee
The result is 
\be \alpha(z) = -\frac{\kappa}{2} \frac{  e^{-\kappa z}}{ \left(e^{2 \kappa  z}+1\right)^{3/2}}\ . \ee
The negative sign out front indicates acceleration to the left, by convention.  

One sees that at late positions, $z\to -\infty$, the acceleration scales like that of the eternally thermal mirror:
\be \alpha(z) \to -\frac{\kappa}{2} e^{-\kappa z}. \ee
Thus the black mirror emits particles with temperature Eq.~(\ref{temperature}) at late positions.  The energy flux and particle spectrum are both exactly the same as black hole radiation for all times \cite{Good:2016oey}.  Qualitatively, this case is like Arcx in Sec~\ref{sec:arcx}, i.e. evaporation never stops as there is an acceleration horizon.  For the black mirror, however, the radiative correspondence is one-to-one to black hole radiation which utilizes the tortoise coordinate \cite{Good:2016oey,Good:2016atu}.

\subsubsection{Drifting Black Mirror} 

The drifting counterpart (Domex)\cite{Good:2016atu} to the black mirror (Omex)\cite{Good:2016oey} does not become asymptotically null, but instead becomes asymptotically drifting, 
\be z(t)=v\,\left(-t- \frac{1}{2\kappa} W[2e^{-2\kappa t}]\right)\ , \ee 
through the inclusion of a multiplicative factor $v < 1$ which is the drifting speed of the mirror at late times, and the maximum speed overall.  The timespace trajectory is the inversion of $z(t)$, 
\be t(z) = - \frac{z}{v} - \frac{1}{\kappa}e^{2\kappa z/v}\ . \ee 

This more simple form can be used to find the proper acceleration as a function of space, 
\be \alpha(z) = -\frac{4 \kappa  v e^{\frac{2 \kappa  z}{v}}}{\left(1-v^2+4 e^{\frac{2 \kappa  z}{v}}+4 e^{\frac{4 \kappa  z}{v}}\right)^{3/2}}\ , \label{eq:domexaz}\ee
where for high drifting speeds,
\be \alpha(z) = -\frac{\kappa  e^{2 \kappa  z}}{2 \left(e^{4 \kappa  z} + e^{2 \kappa  z}\right)^{3/2}} + \mathcal{O(\epsilon)}\ ,\ee
with $v \equiv 1-\epsilon$ and $\epsilon \ll 1$.  

At a late position, where $z$ is a very large negative value, the first  term in the proper acceleration scales as
\be \alpha_1(z) \to -\frac{\kappa}{2} e^{-\kappa z}\ , \ee
demonstrating that for very high speed drifts, the mirror can radiate thermally.  At very late positions, $z\to -\infty$, the full acceleration ultimately drops to zero, $\alpha(z) \to 0$, as seen from Eq.~(\ref{eq:domexaz}), and the mirror eventually settles down to a constant drift. 
Again, like Darcx in Sec \ref{sec:darcx}, this corresponds to an evaporating black hole that dies leaving a remnant \cite{Good:2016atu,Good:2015nja}.

\subsubsection{Null-Self-dual case} \label{sec:selfdual}

A null-self-dual case with maximum speed $v=1$ has trajectory \cite{paper1} 
\be z(t) = -\frac{1}{\kappa}\ln\cosh\kappa t\ , \ee
(again with the asymptotic velocity form approaching the speed of light like Eq.~\ref{eq:zexp}), which becomes 
\be z(\tau) = -\frac{1}{\kappa}\ln\sec\kappa \tau\ , \ee
and acceleration 
\be \alpha(t) = -\kappa \cosh \kappa t\ ,\ee
which becomes 
\be \alpha(\tau) = -\kappa\sec \kappa \tau\ . \ee 

Near $\tau = \pi/2\kappa$, i.e.\ asymptotically in coordinate time, the acceleration scales as
\be \alpha(\tau) = \frac{1}{\tau- \pi/2\kappa} + \mathcal{O}(\tau - \pi/2\kappa)\ . \ee
Again note that using the timespace function $t(z)$, and the Lorentz factor as a function of space, $\gamma(z)$, where $\alpha(z) = \gamma'(z)$, one has the exponential acceleration in space 
\be \alpha(z) = - \kappa e^{-\kappa z}\ , \ee 
and thermal energy flux.

\section{Asymptotes with $1-e^{-2\kp t}$} \label{sec:asymp} 

The next question we consider is whether thermal plateau solutions can be 
made to order, especially given the very few solutions known to date. 
Returning to Eq.~(\ref{eq:zexp}), we can obtain further properties of the 
finite thermal plateau solutions. Recall that we wrote $\dot z=-1+f(t)$ and 
are interested in small $f$. 

When $f=c\,e^{-m\kp t}$ then we find from Eq.~(\ref{eq:fplatf}) 
\be 
F_{\rm plateau}=\frac{m^2\kp^2}{192\pi} \ , 
\ee 
(note that $c$ does not enter here). When $m=2$ we find the conventional 
thermal plateau of Eq.~(\ref{constantenergyflux}). More generally, we can add a 
constant to $f$ to represent the asymptotic drift speed, so 
$f=1-v+c\,e^{-m\kp t}$; this does not change the plateau under the condition 
that $(1-v)\ll e^{-m\kp t}\ll1$. 

We can further establish that after this condition begins to break 
down, as $t$ increases, then the energy flux will cross zero at 
\be 
\kp t_0=-\frac{1}{m}\,\ln\frac{2(1-v)}{c} \ , \label{eq:zeroflux} 
\ee 
and will reach a minimum negative flux dip at 
\be 
\kp t_{\min}=-\frac{1}{m}\,\ln\frac{1-v}{2c} \ . \label{eq:minflux} 
\ee 
Interestingly, the magnitude of the negative dip is 
\be 
F_{\rm min}=-\frac{1}{3}\,F_{\rm plateau} \ . \label{eq:minmax} 
\ee 
As time increases even further, the energy flux approaches zero 
as $(1-v)^{-1} e^{-m\kp t}$. 

The asymptotic drifting case in \cite{paper1} has the correct asymptotic 
form but we did not consider velocities close enough to the speed of 
light, as in Sec.~\ref{sec:selfdual}, to realize the flux plateau on the positive time side. We begin 
with this. 

\begin{figure}[htbp]
\centering 
\includegraphics[width=\columnwidth]{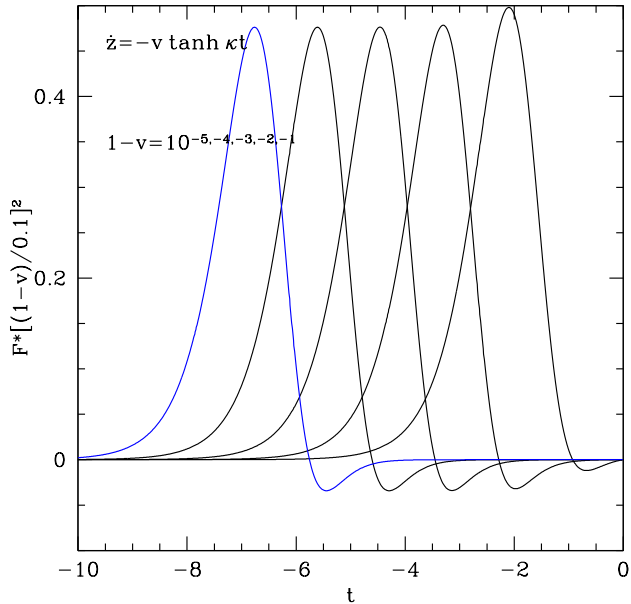} 
\caption{The energy flux to the right of the asymptotically drifting 
self-dual mirror for $t<0$. The different drift speeds range from 
$1-v=10^{-1}$ (rightmost peak) to $10^{-5}$ (leftmost peak, in blue). 
}
\label{fig:driftvneg} 
\end{figure}

The self-dual case asymptotically drifting with maximum speed $v$ 
has $\dot z=-v\tanh \kp t$. The $t<0$ energy flux amplitude scales as 
$(1-v)^{-2}$ and the maximum (positive flux) and minimum (negative 
flux) shift in position as $\Delta t\approx 1.2\log (1-v)$. There 
are only positive peaks and negative dips, with no nonzero plateau, 
as seen in Fig.~\ref{fig:driftvneg}.

\begin{figure}[htbp]
\centering 
\includegraphics[width=\columnwidth]{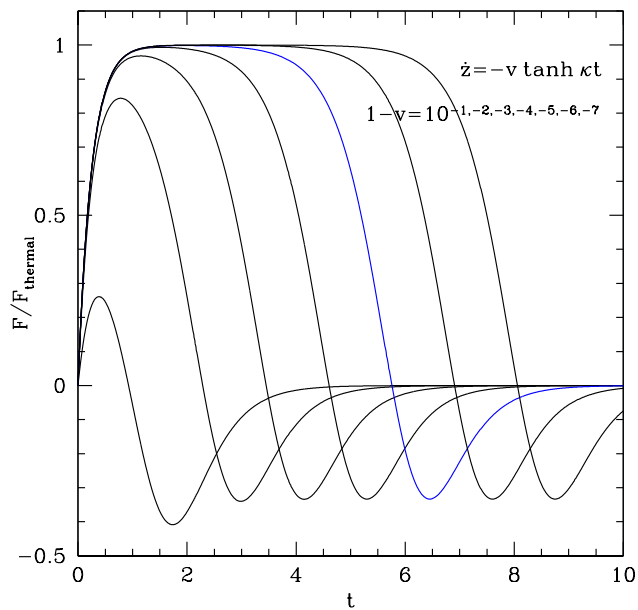} 
\caption{The energy flux to the right of the asymptotically drifting 
self-dual mirror for $t>0$. The different drift speeds range from 
$1-v=10^{-1}$ (leftmost peak) to $10^{-7}$ (rightmost peak), with 
the $1-v=10^{-5}$ case in blue. 
}
\label{fig:driftvpos} 
\end{figure}

However, for $t>0$ the energy flux behavior is quite different, as 
seen in Fig.~\ref{fig:driftvpos}. Now there is a thermal plateau, with 
\be 
F_{\rm max}=\frac{\kp^2}{48\pi} \ , 
\ee 
evident for drift velocities approaching the speed of light, lasting 
from $\kp t\approx 1$ to $\kp t\lesssim\ln(1-v)$, as predicted above. The minimum dips, i.e.\ 
the most negative fluxes have 
\be 
F_{\rm min}=-\frac{1}{3}\,F_{\rm max} \ , 
\ee 
and are shifted in $\kp t$ by $(-1/2)\Delta\ln(1-v)$, as can be 
understood from Eq.~(\ref{eq:minflux}). 
Note that Darcx and Domex follow the same min/max relations, as they 
are particular cases of the asymptotically drifting physics.

\begin{figure}[htbp]
\centering 
\includegraphics[width=\columnwidth]{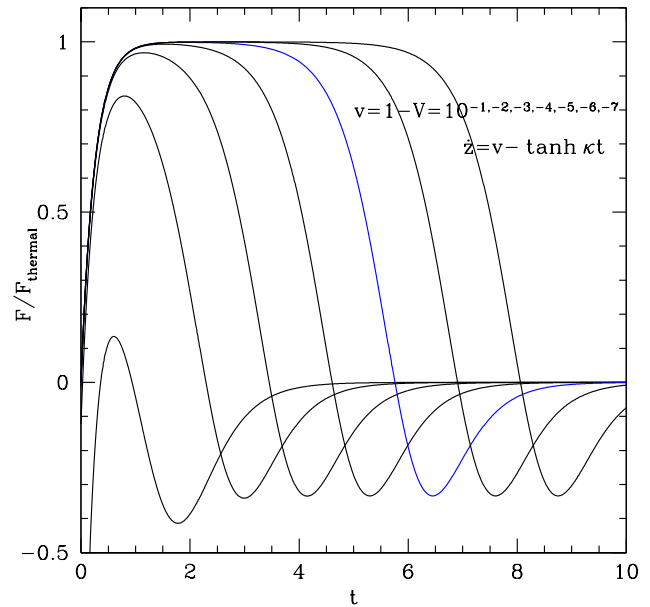} 
\caption{The energy flux to the right of the additively shifted 
asymptotically drifting mirror for $t>0$. The different drift speeds 
range from $v=1-V=10^{-1}$ (leftmost peak) to $10^{-7}$ (rightmost peak), with 
the $v=1-V=10^{-5}$ case in blue. The flux approaches that of the 
multiplicative case for high velocities $V$ and late times. 
}
\label{fig:driftadd} 
\end{figure}

Recall that the asymptotically drifting self-dual mirror was a 
multiplicative shift of the $\dot z=-\tanh\,\kp t$ case. As described 
in \cite{paper1} we can also 
investigate an additive shift of this case, such that 
\be 
\dot z=v-\tanh\,\kp t \ . 
\ee 
This is no longer self dual, but is asymptotically drifting, with 
speed $\dot z\to -1+v$, rather than $-v$ in the multiplicative case. 
This additive case also has an energy flux plateau for $t>0$, as seen 
in Fig.~\ref{fig:driftadd}.

\begin{figure}[htbp]
\centering 
\includegraphics[width=\columnwidth]{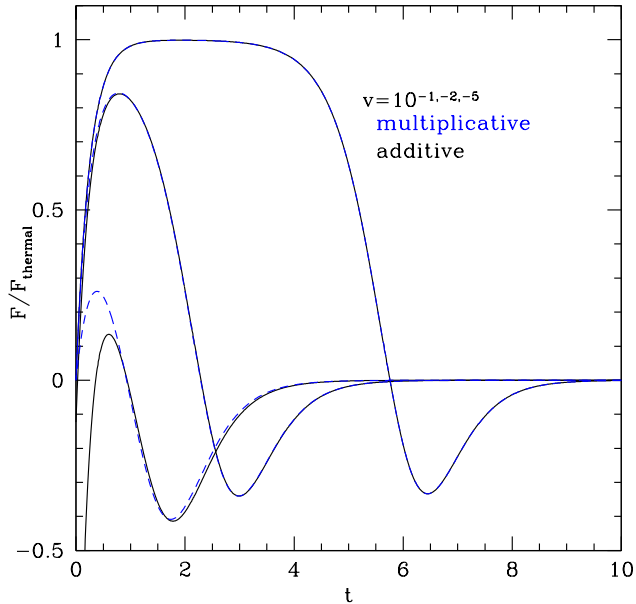} 
\caption{The energy fluxes of the additively shifted and multiplicatively 
shifted asymptotically drifting mirror for $t>0$ are quite similar. 
We show drift speeds of $v=1-V=10^{-1}$, $10^{-2}$, and $10^{-5}$, from 
left to right. Note that while the multiplicative case has zero flux, 
and $\dot z=0$, at $t=0$, the additive case does not. 
}
\label{fig:comparedrift} 
\end{figure}

When the drift speed moves away from the speed of light, differences 
appear between the energy fluxes from the multiplicative and additive 
cases. We compare the two\footnote{ 
For this comparison only, we will refer to the multiplicative shift 
as $V$. The asymptotic drift speed goes as $1-v$ in the additive case, and 
so we look at small $v$, and goes as $V$ in the multiplicative case, so 
we look at $V\approx1$. In either case the important physical aspect 
is that the drift speed is near the speed of light.} 
in Fig.~\ref{fig:comparedrift}. Differences 
arise for larger values of $v$ (smaller values of $V$), 
at $\kp t\lesssim1$. Recall that at 
$t=0$, the multiplicative case has zero velocity while the additive 
case has finite velocity. This corresponds to different formation 
scenarios in the analog black hole case.

\begin{figure}[htbp]
\centering 
\includegraphics[width=\columnwidth]{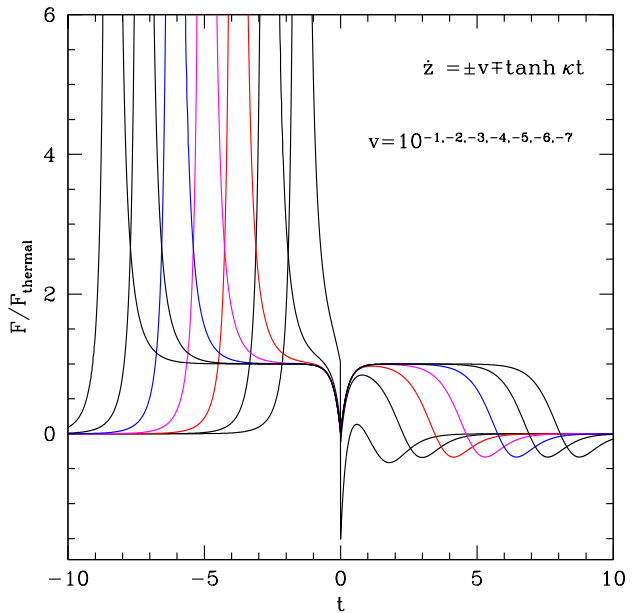} 
\caption{The energy flux to the right of the additively shifted 
self-dual mirror for all times. Note the thermal plateaus for 
both $t<0$ and $t>0$. The different drift speeds 
range from $1-10^{-1}$ (innermost peak) to $1-10^{-7}$ (outermost peak), 
with the $1-10^{-3,-4,-5}$ cases in red, magenta, blue respectively. 
}
\label{fig:driftaddall} 
\end{figure}

We can extend the additive case to $t<0$ in one of two ways. (If we 
do nothing, then note that $|\dot z|>1$ can occur for $t<0$.) First, 
we could use $|t|$ rather than $t$ in the argument of tanh. If we do 
this, we find the energy flux for $t<0$ does not exhibit a plateau, 
same as the multiplicative case. Second, we could use $|t|$ but also 
allow the overall sign of $\dot z$ to flip. This second approach 
corresponds to the trajectory 
\be 
z=v|t|-\kp^{-1}\,\ln\,[\cosh \kp |t|] \ . 
\ee 
Note that this is even in time and so is a self-dual mirror. We then 
find that this solution exhibits an energy flux thermal plateau for 
$t<0$ as well as $t>0$. Figure~\ref{fig:driftaddall} illustrates 
the results.

The thermal plateaus are at the same temperature, given by the 
acceleration parameter $\kp$. On the $t<0$ side there is a strong 
positive energy flux peak before the plateau, while on the $t>0$ side 
there is a negative flux dip after the plateau. The asymptotic drift 
speed goes as $1-v$, so again this case exhibits plateaus when $v$ is 
very small (so the drift speed is near the speed of light). 

\begin{figure}[htbp]
\centering 
\includegraphics[width=\columnwidth]{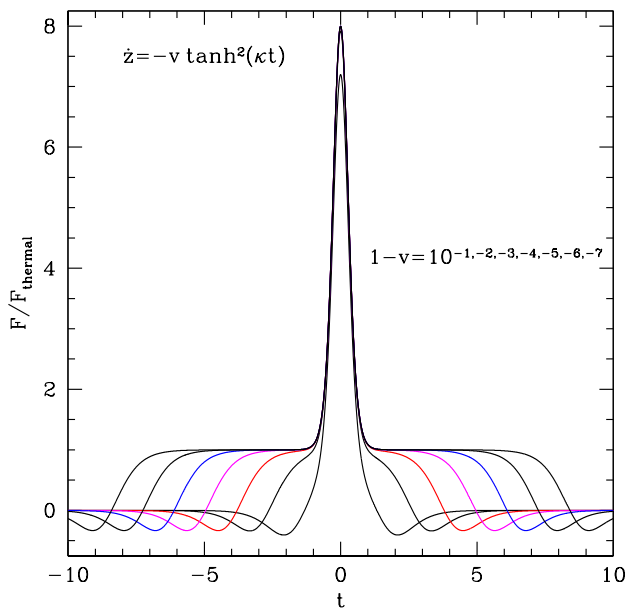} 
\caption{The energy flux to the right of the time symmetric, asymptotically 
drifting mirror for all times. Note the thermal plateaus for 
both $t<0$ and $t>0$. The different drift speeds 
range from $1-10^{-1}$ (innermost curve) to $1-10^{-7}$ (outermost curve), 
with the $1-10^{-3,-4,-5}$ cases in red, magenta, blue respectively. 
}
\label{fig:vtanh2} 
\end{figure}

A more elegant method of attaining thermal plateaus for negative and 
positive time is to square the tanh function. Taking 
\be 
\dot z=-v\,\tanh^2 \kp t \ , \label{eq:vtanh2} 
\ee 
we have an asymptotically drifting mirror with symmetric thermal 
phases and with no discontinuities at $t=0$. The mirror trajectory 
is $z=-vt+(v/\kp)\,\tanh \kp t$. The energy flux is plotted in 
Fig.~\ref{fig:vtanh2}.

\begin{figure}[htbp]
\centering 
\includegraphics[width=\columnwidth]{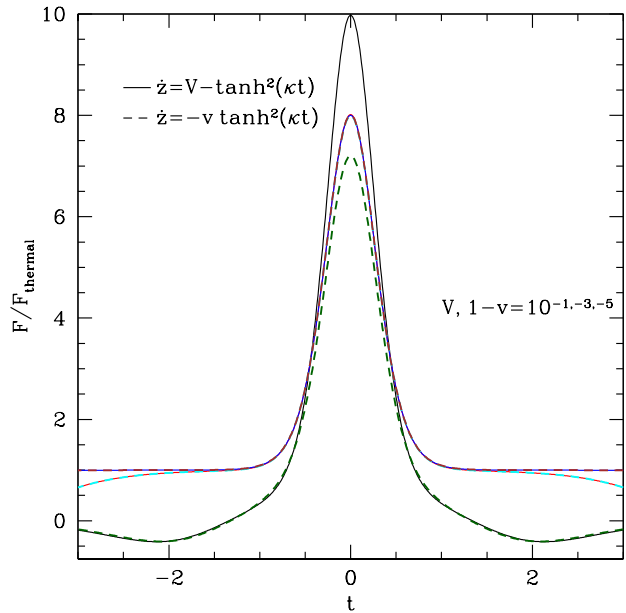} 
\caption{The energy fluxes of the additively shifted (solid) and multiplicatively shifted (dashed) $\tanh^2$ mirrors are highly similar except for $\kp t\ll1$ and drift speeds ($1-V$ in the additive case, $v$ in the multiplicative case) not very close to the speed of light. We show drift speeds of $V=1-v=10^{-1}$, $10^{-3}$, and $10^{-5}$, from inner to outer curves. 
}
\label{fig:comparevtanh2} 
\end{figure}

With the additive shift of $\tanh^2 \kp t$ rather than the multiplicative shift of Eq.~(\ref{eq:vtanh2}), that is, 
\be 
\dot z = V-\tanh^2 \kp t \ , 
\ee 
we again find symmetric thermal plateaus. The energy flux is very similar to Fig.~\ref{fig:vtanh2}, so we only show a comparison of multiplicative and additive cases in the $\kp t\ll1$ region where they may differ, in Fig.~\ref{fig:comparevtanh2}. 
Again, this corresponds to different black hole formation histories.

\begin{figure}[htbp]
\centering 
\includegraphics[width=\columnwidth]{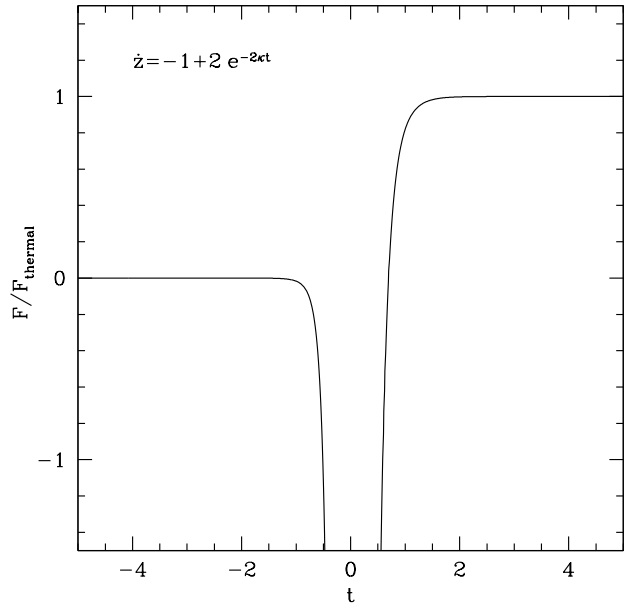} 
\caption{Taking the asymptotic mirror velocity behavior, $\dot z=-1+2e^{-2\kp t}$, to hold for all times yields a (half-)eternal thermal state. 
}
\label{fig:expall} 
\end{figure}

One can also keep the asymptotic form for all times, i.e.\ $\dot z=-1+2e^{-2\kp t}$. Figure~\ref{fig:expall} shows that this gives  a plateau for all $\kp t\gg1$, out to infinity, equivalent to an 
eternal black hole. If one was concerned with small $t$ one would 
want to adjust the form there as the acceleration is infinite at $t=0$. For $t<0$ with $\kp |t|\gg1$, the flux vanishes. One could multiplicatively shift this, and use $|t|$ in the exponential, yielding plateaus for both negative and positive times, and then the acceleration is finite but discontinuous at $t=0$. Alternately one could make the prefactor $\mp v$, which regularizes the acceleration at $t=0$ but makes the speed discontinuous there.

\section{Bespoke Thermal Flux Plateaus} \label{sec:bespoke} 

We have seen that various functional forms can give a thermal energy flux 
plateau. Here we examine whether this can be made on demand. All of 
these functions, and the Carlitz-Willey solution as well, can be written as an 
infinite series expansion. Just as in \cite{paper1} we studied how going 
from a finite to infinite series took the self-dual asymptotic 
static mirror to the drifting one, we can investigate here various 
series expansions and their relation to a finite vs infinite thermal 
flux plateau. 

The conditions for a thermal flux plateau are that the flux is 
thermal, and unchanging at some order, i.e.\ 
\bea 
F&=&\frac{\kap^2}{48\pi}\ , \\ 
\dot F|_{\tilde t}&\approx&0\ , 
\eea 
where the derivative is evaluated at some plateau time $\tilde t$. 

We consider two expansions, one around $t=0$ and one around $t=\infty$, and seek to build up thermal flux plateaus extending 
from small $t$ onward and from large $t$ inward. The small and 
large $t$ expansions, respectively, are 
\bea 
\dot z&=&\sum a_n t^n\ , \\
\dot z&=&\sum a_n\,e^{-2nt}\ , 
\eea 
where $t$ is measured in units of $\kap^{-1}$. We of course need 
to check that the amplitude of $\dot z$ never exceeds unity.

\subsection{Small time expansion} \label{sec:small}

For the small $t$ expansion, calculating the flux at each order 
implies that the constant flux term, i.e.\ the one proportional to $t^0$, is 
\be 
12\pi F^{[0]}=\frac{-2a_2(1-a_0^2)-3a_0a_1^2}{(1-a_0)^4(1+a_0)^2}\ . 
\ee 
Thermality then imposes the constraint 
\be 
a_2=\frac{-3a_0 a_1^2}{2(1-a_0^2)}-\frac{1}{8}(1-a_0)^3(1+a_0)\ . 
\ee 
At the next order, imposing that the time variation vanishes, i.e.\ 
$F^{[1]}$ proportional to $t^1$ goes to 0, one obtains a constraint $a_3(a_0,a_1,a_2)=a_3(a_0,a_1)$. This continues for all orders, so 
one has a two parameter family of mirror solutions with thermal 
flux plateaus, given by $(a_0,a_1)$. 

The physical meanings for the parameters are that $\dot z(t=0)=a_0$, 
and $\al(0)=a_1/(1-a_0^2)^{3/2}$. If one chooses 
$a_1=0$, i.e.\ no acceleration at $t=0$, then all odd terms in 
the expansion vanish. 

While one might expect that the expansion, 
and hence thermal plateau, is only good for $t\ll1$, in fact it 
can last for considerably longer. This can be seen analytically 
in the $a_1=0$ case, where one finds that for $a_0=1-\eps$, 
with $\eps\ll1$, i.e.\ a velocity near the speed of light, 
the even terms in the expansion are of order $\eps^{n+1}t^n$. 
Therefore the expansion, and plateau, are valid for 
\be 
t\ll\eps^{-(n+1)/n} \ . \label{eq:breakt}
\ee 
If $\eps\sim10^{-2}$, say, i.e.\ the mirror velocity at $t=0$ is 
$v=a_0>0.99$, then the plateau lasts for $t\approx[0,30]$, so for many 
characteristic time scales $\kap^{-1}$. Beyond this, the model will 
break down for a finite number of terms in the expansion, with eventually 
$\dot z>1$. 

\begin{figure}[htbp]
\centering 
\includegraphics[width=\columnwidth]{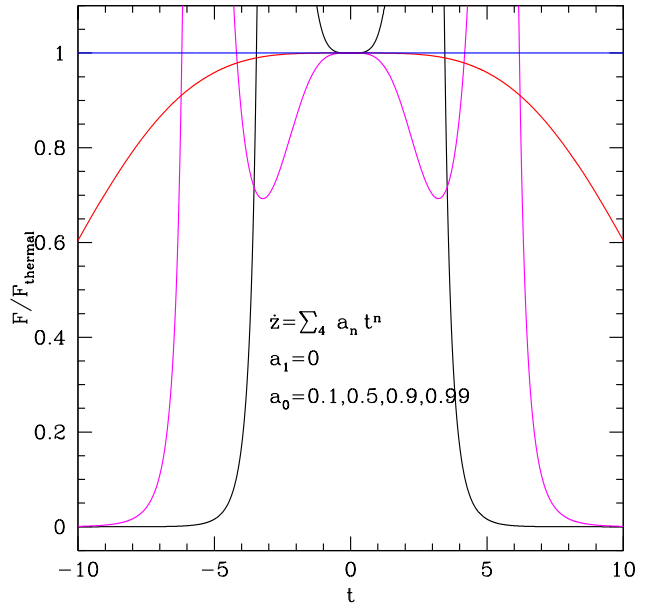} 
\caption{Taking the mirror velocity as a series expansion at small times, 
we can create a family of solutions with thermal plateaus 
extending for a finite time. When $a_0$, the velocity at $t=0$, 
approaches the speed of light, the plateau can last to $t\gg1$ 
($a_0=0.1, 0.5, 0.9, 0.99$ are the black, magenta, red, blue curves). 
}
\label{fig:texp} 
\end{figure}

Figure~\ref{fig:texp} illustrates the behaviors, using an expansion 
up to $n=4$. A long thermal flux plateau can be readily constructed. 
Note that for the choice of $a_1=0$ shown, the flux is symmetric 
in time as the mirror is self dual. As the mirror velocity at $t=0$ approaches 
the speed of light, 
the plateau lasts longer, until it eventually violates the condition 
(\ref{eq:breakt}). Beyond that, where the model is invalid, the flux 
can diverge before eventually going to zero. 
Such finite plateau solutions can be used to 
study the onset at least of black hole evaporation.

One particularly interesting case is that with zero velocity at 
$t=0$, i.e.\ with $a_0=0$ as well as $a_1=0$. This takes the form 
\be 
\dot z=-\frac{1}{8}t^2-\frac{1}{192}t^4+{\mathcal O}(t^6) \ , 
\ee 
and exhibits a thermal plateau for $t<1$. At large times the flux 
drops to zero. Of note is that this 
mirror solution has no negative energy flux at any time.\footnote{A negative Schwarzian derivative of $p(u)$ corresponds to positive energy emission of the black hole-moving mirror.  The operation is invariant under more than just inertial transformations: linear fractional transformations, i.e.\  M{\"o}bius functions \cite{Fabbri:2005mw}, have vanishing energy flux. }

With an infinite series, the expansion can become a function like 
tanh or the product log $W$, giving familiar results with thermal 
plateaus and being well behaved at all times. Alternately, one can 
add a graceful exit from the series expansion to avoid divergences. 
The main point is that we have constructed a well-defined two 
parameter family exhibiting a finite period of thermal solutions for evaporating black holes and radiating moving mirrors.

\subsection{Large time expansion}

In the large $t$ case, we look for a semi-infinite period of thermal 
flux, with a plateau extending from $t=\infty$ inward, potentially 
to $t\approx\kap^{-1}$ with a finite series expansion. We can write the 
series as 
\be 
\dot z=\sum a_n q^n\ , 
\ee 
where $q=e^{-2\kap t}$. Note that at $t=\infty$ (we only consider 
positive $t$ here), $\dot z(\infty)=a_0$. To get a thermal plateau 
it is natural to take $a_0=-1$, as in the earlier solutions we 
studied such as $\dot z=-1+2e^{-2\kap t}$. 

We find that in this case the $q^0$ (time independent) term of the flux gives 
thermality at $t=\infty$ for any choices of the other $a_n$, with $a_1\ne0$. 
Moreover, the $q^1$ term vanishes, so we have a two parameter 
family of semi-infinite thermal flux solutions, determined by 
the values of $(a_1,a_2)$. At second order, the constraint condition 
becomes 
\be 
a_3=-\frac{3}{8}\,\left(a_1^3+2a_1 a_2-2\frac{a_2^2}{a_1}\right)\ . 
\ee 
That is, the thermal plateau condition is given by $a_3(a_1,a_2)$, 
and so on for higher order terms in the series. 

One particular realization of this condition would be $a_1=2$, 
$a_2=-4$, which imposes $a_3=-9$. This choice corresponds to the 
series coefficients for the Carlitz-Willey solution, through third order. 
However, as we said there is actually an entire two parameter 
family of such solutions, rather than just this one, which we could solve order by order to 
get a thermal plateau. 

\begin{figure}[htbp]
\centering 
\includegraphics[width=\columnwidth]{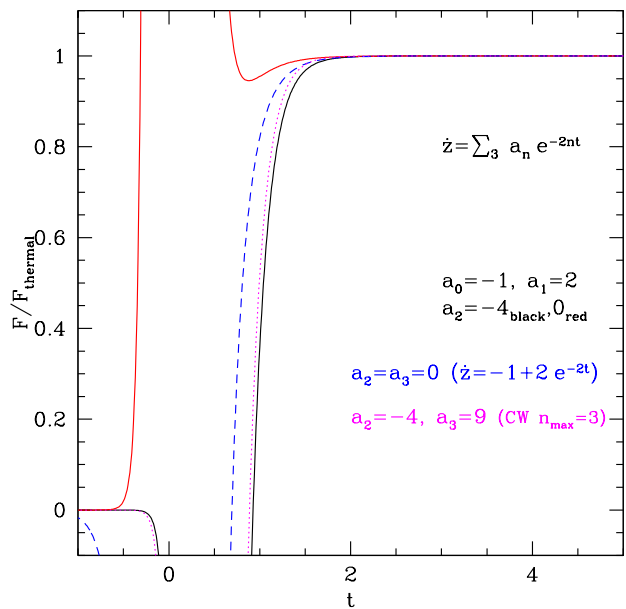} 
\caption{Taking the mirror velocity as a series expansion at large times, 
we can create a family of solutions with thermal plateaus 
extending for a semi-infinite time. For this figure, we only vary 
$a_2$, keeping the other parameter $a_1$ fixed, yet still obtain 
a variety of thermal solutions. When we neglect all terms beyond first order, 
we have the previous solution of $\dot z=-1+2\,e^{-2\kap t}$ (dashed blue curve), 
but we see that inclusion of the further terms in the series can 
improve the solution. Even when we keep $a_2=0$, using the $a_3$ 
determined by the thermality condition (red curve) lengthens the 
plateau. The first three terms of the Carlitz-Willey (CW) expansion give 
the magenta dotted curve, similar to the CW eternally thermal 
solution corresponding to an infinite series. The solid black curve 
with that same $a_2$ but no third order term is also quite similar. 
}
\label{fig:qexp} 
\end{figure}

Figure~\ref{fig:qexp} illustrates some mirror solutions with these 
properties. Several different choices of $a_2$ lead to the same 
qualitative behavior. In particular, note the thermal plateau 
extends over $t\approx[1,\infty]$. Such solutions can be used to 
study eternal black holes.

Again, with an infinite series, the expansion can become a function like 
$\tanh^2$ or the product log $W$ (as Carlitz-Willey is), avoiding the 
flux divergence where the expansion breaks down at $q\sim1$, 
i.e.\ $t\lesssim1$, and giving familiar results with thermal 
plateaus and being well behaved at all times. Alternately, one can 
add a graceful exit from the series expansion to avoid divergences (e.g.\ a smooth match onto a static mirror).  
The main point is that we have constructed a well defined two 
parameter family exhibiting semi-infinite thermal solutions, possible forming and then eternal black hole analogs.

\subsection{Constant asymptotic acceleration}

The series expansion solutions have infinite acceleration at large 
$t$ (for a finite series). It is interesting to study a solution 
with a large time flux plateau but constant finite asymptotic acceleration. 
Asymptotically constant acceleration means the mirror approaches the speed of light as $\dot z\to 1-\mathcal{O}(t^{-2})$, unlike the exponential approach. 
The linear-square root (LS) family has this property, with 
\be 
\dot z=\left(\frac{t}{\sqrt{t^2+1}}\right)^n \ , \label{eq:lsn}
\ee 
giving $\al(t\to\infty)\to 1/\sqrt{n}$. 

At $t=\infty$ (and $t=-\infty$ for $n$ even), the flux reaches a 
plateau with 
\be 
12\pi F(t\to\infty)\to \frac{2(n-1)}{n^2} \ . 
\ee 
Note that $F$ only equals the amplitude of the usual thermal plateau value of 
$1/(48\pi)$ for $n=4\pm2\sqrt{2}$, and it is not known if the particle emission has a thermal spectrum from this power-law rather than exponential asymptote. 

\begin{figure}[htbp]
\centering 
\includegraphics[width=\columnwidth]{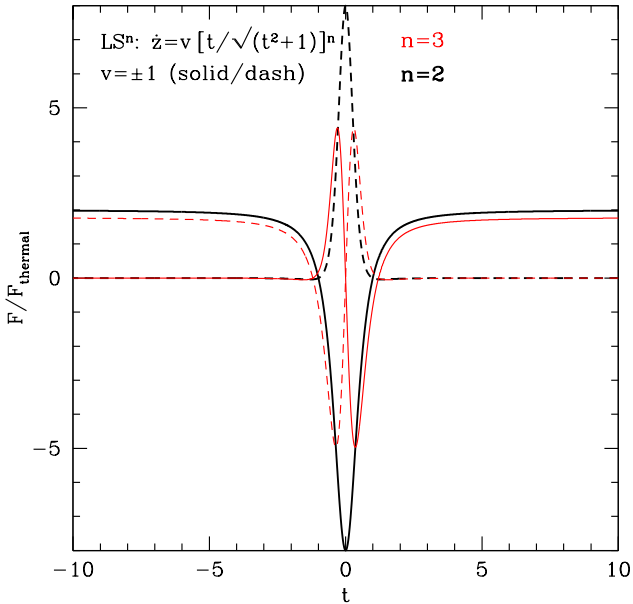} 
\caption{For the $LS^n$ mirror solution, semi-infinite or almost 
eternal thermal flux plateaus exist at large $t$, although at a 
modified temperature. 
}
\label{fig:lsn} 
\end{figure}

Figure~\ref{fig:lsn} illustrates the flux behavior for $n=2$ and 
3. For $n$ even, the flux is symmetric about $t=0$, and a plateau exists if the maximum mirror velocity is 1, as in Eq.~(\ref{eq:lsn}). 
If we switch the asymptotic sign of $\dot z$ then there is no 
plateau. For $n$ odd, there is a plateau at $t=\pm\infty$ for 
$\dot z\to\pm 1$. Again this solution represents an eternal black 
hole, but with a renormalized surface gravity, i.e.\ temperature. 
No divergences appear in this solution. For $n=3$, despite the radiation continuing at a constant rate for $t\to+\infty$ in coordinate time $t$, the total evaporation energy, $\int F(t)(1-\dot{z})dt$, is finite and analytic as measured by an observer measuring with clock $u$ at future null-infinity, $\mathcal{I}^{R+}$.

\section{Conclusion} \label{sec:concl} 

Thermal energy emission from black holes, i.e.\ the existence of a 
temperature related to black hole mass and surface gravity, is an 
extraordinary phenomenon 
with deep ties to fundamental physics, including both gravitation and 
quantum physics, that are not yet completely understood. 
We have turned to the analogous situation of accelerating mirrors and 
investigated the precise conditions for their thermal energy emission. 

In particular, we have shown that eternal thermal emission is not 
unique to only the one example previously known, the Carlitz-Willey 
mirror trajectory. By exploring the energy flux in terms of proper 
time $\tau$ and retarded time $u$ we found extremely simple expressions 
of energy flux dynamics: $12\pi F(\tau) = - \alpha'e^{2\eta}$ and 
$12\pi F(u) = -\alpha' e^{\eta}$ in the two variables respectively. 
This led to three possible eternally thermal solutions. 

In the simplest case, a thermal energy flux plateau arises from scale 
independence of the acceleration in proper time, $\alpha = \tau^{-1}$. 
This allows rewriting of the Carlitz-Willey case in much simpler form. 
We have also related this to an exponential approach of the mirror 
speed to the speed of light, and connected this to black hole horizons. 

Using the exponential approach to the speed of light 
asymptotic form, we study a number of mirror trajectories to bring out 
both general relations such as when thermal emission breaks down and 
its relation to an ensuing negative energy flux bounce, and the 
distinctions between specific cases, for which we present the energy 
flux as a function of time. Using series expansions at small and at 
large time, we create custom made thermal energy solutions that can 
represent black holes with differing formation and evaporation 
histories, as well as eternal black holes. 

Note that our results require that testing thermal emission from forthcoming laboratory experimental setups for radiation from moving mirrors, e.g.\ by accelerating plasma walls \cite{plasma} or in resonant circuits \cite{wilson}, must be able to achieve exponential accelerating conditions. 

These results for thermal energy emission are not only important in 
their own right as a ``toolkit'' for studying black hole evolution, 
but because of the close relation of energy flux with entanglement 
entropy. Future work will address this avenue for understanding 
the information content of black holes and mirrors \cite{Hotta:2015yla,Chen:2017lum}, and the relation of the 
positive-only energy flux solution we have found (see Sec.~\ref{sec:small}) 
to properties such as minimum black hole mass.

\acknowledgments 
MG thanks Benito A.\ Ju{\'a}rez-Aubry, Pisin Chen, Yen Chen Ong and Robert Caldwell for stimulating discussions; EL thanks Robert Caldwell and Misao Sasaki. MG was funded in part from the Julian Schwinger Foundation under Grant 15-07-0000 and the ORAU and Social Policy grants at Nazarbayev University. EL is supported in part by the Energetic Cosmos Laboratory and by 
the U.S.\ Department of Energy, Office of Science, Office of High Energy 
Physics, under Award DE-SC-0007867 and contract no.\ DE-AC02-05CH11231. 

\appendix

\section{Eternal Thermality: Carlitz-Willey}

The Carlitz-Willey \cite{Carlitz:1986nh} mirror solution trajectory, $z(t)$, \cite{Good:2012cp,Good:2013lca} is known to give eternally thermal energy flux. 
Here we uncover its relation to the results in the main text.

\subsection{Exponential Acceleration in Null Time is Thermal} 

Beginning with the Carlitz-Willey mirror trajectory \cite{Good:2012cp,Good:2013lca},
\be z(t) = -t - \frac{1}{\kp}W(e^{-2\kappa t})\ ,\ee 
consider the acceleration in terms of the null time $u$. This can be found 
by substitution into $v(t) = t + z(t)$.  Inversion gives $t(v)$, for $v<0$,
\be t(v) = -\frac{\ln \left(-\kappa  v e^{-\kappa v}\right)}{2 \kappa }\ .\ee
Using this in 
\be \alpha(t)= -\frac{\kappa}{2\sqrt{W(e^{-2 \kp t})}} \ , \ee
gives, with $v<0$, 
\be \alpha(v) = -\frac{1}{2}\sqrt{\frac{\kappa}{-v}}\ . \ee
The trajectory, as expressed in ray-tracing functions,
\be v = p(u) = -\frac{1}{\kappa}e^{-\kappa u}\ , \ee
allows substitution into $\alpha(v)$ to obtain $\alpha(u)$,
\be \alpha(u) = -\frac{\kappa}{2}\, e^{\kappa u/2}\ . \ee
Therefore, we see that the Lorentz invariant proper acceleration for the 
Carlitz-Willey eternally thermal solution can be written simply as  exponential acceleration as 
measured by null time $u$, rather than the relatively more complicated product log function 
as measured in coordinate time $t$.

\subsection{Scale Independent Acceleration in Proper Time is Thermal} \label{sec:apxcw} 

An even greater simplification occurs in terms of proper time $\tau$. 
Again starting with the eternally thermal Carlitz-Willey trajectory 
\be \label{cwtrajectory} z(t) = -t - \frac{1}{\kp}W(e^{-2\kappa t})\ ,\ee
substitution into $\alpha(t) = \gamma^3(t) \ddot{z}(t)$ gives the proper acceleration,
\be \label{thermalacc} \alpha(t)= -\frac{\kappa}{2\sqrt{W(e^{-2 \kp t})}} \ . \ee
In stark contrast to the thermal emission of the Unruh effect, it is notable that the proper acceleration is not constant, despite the constant energy flux.  One way to express this in terms of proper time is to integrate 
from coordinate time to proper time,
\be \tau(t) = \int \frac{dt}{\gamma(t)}\ . \ee
This integral gives 
\be \label{taut} \tau(t) = -\frac{2}{\kappa} \sqrt{W(e^{-2 \kp t})}\ , \ee
the reciprocal of Eq.~(\ref{thermalacc}); therefore 
\be \alpha(\tau) = \tau^{-1}\ ,  \ee
for all time $\tau < 0$. For early times $\tau \to -\infty$, the mirror is asymptotically inertial, 
$\alpha \to 0$.  As $\tau \to 0$, the acceleration of the mirror asymptotically diverges, 
$\alpha \to -\infty$. A key result is that the acceleration, $\alpha(\tau)$, is independent 
of  scale, e.g.\ $\kappa$. 

One can carry out the same analysis in terms of rapidity. 
Inverting Eq.~(\ref{taut}),
\be t(\tau) = -\frac{1}{\kappa}\ln\frac{\kappa \tau}{2} - \frac{\kappa \tau^2}{8} \ ,\ee
and expressing the trajectory $z(t)$, Eq.~(\ref{cwtrajectory}), in terms of proper time, $\tau$,
\be z(\tau) = \frac{1}{\kappa}\ln\frac{\kappa \tau}{2} - \frac{\kappa \tau^2}{8}\ . \ee
Finding the proper velocity (celerity) is straightforward: 
\be w(\tau) \equiv \frac{dz(\tau)}{d\tau} = \frac{1}{\kappa \tau}- \frac{\kappa \tau}{4}\ , \ee
and the rapidity is the $\textrm{arcsinh}$ of the celerity,
\be \eta(\tau) = \textrm{arcsinh}\left(\frac{1}{\kappa \tau}- \frac{\kappa \tau}{4}\right)\ .\ee
Finally, the proper acceleration is found by taking the proper time derivative of the rapidity,
\be \alpha(\tau) \equiv \frac{d\eta(\tau)}{d\tau} = \tau^{-1}\ . \ee

\subsection{Summary: Eternally Thermal Acceleration}

The following equations provide a precis of the proper acceleration of the eternally thermal moving mirror of Carlitz-Willey.   

Proper time (note scale independence):
\be \alpha(\tau) = \tau^{-1}, \quad -\infty < \tau < 0 \ee
Null coordinates $u = t-z$ and $v = t+z$: 
\be \alpha(u) = -\frac{\kappa}{2}\, e^{\kappa u/2}\ , \quad -\infty < u < \infty, \ee
\be \alpha(v) = -\frac{1}{2} \sqrt{\frac{\kappa}{|v|}}\ , \quad -\infty < v < 0, \ee
Spacetime coordinates $(z,t)$: 
\be \alpha(t)= -\frac{\kappa}{2\sqrt{W(e^{-2 \kp t})}}, \quad -\infty < t < \infty, \ee
\be \alpha(z) = \pm\frac{\kappa}{2\sqrt{-W(-e^{2 \kp z})}}, \quad -\infty < z < -\frac{1}{2\kappa}\ . \ee
The conventional expression in terms of the product log obscures the fact 
that at late times, the acceleration scales like 
\be  t\to+\infty, \quad \alpha(t) \to -\frac{\kappa}{2} e^{\kp t}\ , \ee
and, because the mirror accelerates off to the left by convention,  
\be z\to-\infty, \quad \alpha(z) \to -\frac{\kappa}{2} e^{-\kp z}\ . \ee

\section{Rapidity and Acceleration} \label{sec:apxtsfm} 

For convenience, we here summarize expressions involving the velocity, rapidity, 
and acceleration. The rapidity is defined in terms of the proper velocity (celerity) $w$, the Lorentz factor $\gamma$, and the velocity $v$ respectively as 
\be \eta = \sinh^{-1}{w} = \cosh^{-1}\gamma = \tanh^{-1}{v}\ , \ee
while the proper acceleration in terms of rapidity is 
\be \alpha = \frac{d}{d\tau}\eta = e^{-\eta}\frac{d}{du}\eta = \gamma \frac{d}{dt}\eta = e^\eta \frac{d}{dv}\eta = w\frac{d}{dx}\eta\ , \ee
or more compactly, 
\be \alpha = \frac{d}{d\tau}\eta= \frac{d}{dt} w = \frac{d}{dx}\gamma\ . \ee
And as usual 
\be \alpha = \gamma^3 \frac{d}{dt} v\ . \ee


\end{document}